\documentclass[preprint,showpacs,pra,amsmath,amssymb]{revtex4-1}
\usepackage{bm}
\begin{document}

\title{Quantum adiabatic theorem in light of the Marzlin-Sanders inconsistency}
\author{Juan Ortigoso}

\affiliation{Instituto de Estructura de la Materia, CSIC, Serrano 121, 28006 
Madrid, Spain}

\date{26 September 2012}

\begin{abstract}
A  consensus that questions the perfunctory use of the quantum adiabatic theorem has emerged since Marzlin and Sanders [Phys. Rev. Lett. {\bf 93}, 160408 (2004)] showed the existence of an inconsistency in the applicability of the theorem. Further analysis proved that the inconsistency may arise from the existence of resonant terms in the Hamiltonian, but recent work indicates that the debate about the full extent of the problem remains open. Here,  we first show that key premises required in the standard demonstration of the theorem do not hold for a dual Hamiltonian involved in the Marzlin-Sanders inconsistency. Also, we show that two simple conditions can identify systems for which the adiabatic approximation fails, in spite of satisfying traditional quantitative conditions that were believed to guarantee its validity.  Finally, we prove that the inconsistency only arises for Hamiltonians that contain resonant terms whose amplitudes go asymptotically to zero.
\end{abstract}

\pacs{03.65.Ca, 03.65.Ta, 03.65.Vf}

\maketitle

\section{Introduction}
\label{uno}

A folk quantum adiabatic theorem (QAT) establishes that a system initially described by an eigenstate associated with an isolated eigenvalue, of an instantaneous time-dependent Hamiltonian, $H(0)$,  will be at further times in the corresponding eigenstate of $H(t)$, if $H$ changes slowly enough. 
The rate at which the adiabatic limit is reached depends on the gap separating the eigenvalue of interest from the rest of the energy spectrum.  
 
The theorem was initially proven by Born and Fock \cite{byf} for bounded Hamiltonians with discrete energy levels and extended later by Kato \cite{kato} who removed the assumption of spectral discreteness, provided the initial eigenstate corresponds to a discrete eigenvalue. On the other hand,  Avron and Elgart \cite{avron} showed that an adiabatic theorem can be formulated for systems with no spectral gaps. In practice,  when the Hamiltonian variation is not infinitely slow,  the adiabatic theorem does not hold although, under certain conditions, the dynamics can be approximately adiabatic.  Statements about these conditions constitute the so-called adiabatic approximation \cite{mackenzei}. The present paper deals, somehow simultaneously, with the theorem and the approximation, but we emphasize that both concepts are not synonymous and they must be clearly distinguished.

The publication in 2004 of a paper \cite{marzlin} pointing out an inconsistency in the theorem has resulted in a new consensus among an increasing number of physicists, which indicates that ``extreme caution should be used when interpreting results based on the standard application of the theorem" \cite{zoller}. The so-called 
Marzlin-Sanders inconsistency \cite{marzlin} can be presented in the following simple way due to Tong {\it et al.} \cite{tong}: If a system $S_a$ evolves adiabatically under the dynamics generated by a Hamiltonian  $H_{a}(t)$, a dual system $S_b$, described by the Hamiltonian  $H_{b}(t)=-U_{a}(t,0)^{\dagger}H_{a}(t)U_{a}(t,0)$, where $U_{a}(t,0)$ is the exact evolution operator for system $S_{a}$, does not evolve adiabatically in general. 
However, both systems satisfy the same quantitative requirements

\begin{equation}
\left|\frac{\langle E_{m}(t)|\dot{E}_{n}(t)\rangle}{E_{n}(t)-E_{m}(t)}\right|<<1, \;\;\;\;\; m\ne n \;\;, \forall t\;,
\label{Eq1}
\end{equation}

\noindent 
that traditionally were considered to guarantee that a given Hamiltonian (with eigenpairs $E_{n}(t), |E_{n}(t)\rangle$) evolves slowly enough.

Tong {\it et al.} \cite{tong} attributed the inconsistency to the
insufficiency of Eq. (\ref{Eq1}). Comparat \cite{comparat} argued that Eq. (\ref{Eq1}) is valid to detect adiabatic systems except when the Hamiltonian contains oscillatory terms. 
MacKenzie {\it et al.} \cite{mackenzei} emphasized the differences between adiabatic approximation and adiabatic theorem, and explained that the inadequacy of the quantitative conditions arises in situations where the approximation, but not the theorem, is invalid.  Duki {\it et al.} \cite{deny0} showed that the paradox results from the breakdown of the adiabatic approximation for the dual system. Similarly, Amin \cite{explanations} showed that the inconsistency arises because the dual Hamiltonian contains resonant terms. Other authors \cite{vertesi}, \cite{jansen} indicated that the dual Hamiltonian, $H_{b}$, present multiple time scales and for that reason adiabatic theorems do not apply to it. Finally, some works have denied the inconsistency \cite{deny1,deny2},

The current consensus indicates that QAT  has been rigourously proven beyond any doubt, and it is not in question. In fact, no problems have been reported so far in the derivations of the theorem.
The dominant notion seems to be that the Marzlin-Sanders paradox is related to a failure or insufficiency of the conditions that assure that the adiabatic approximation is applicable to a particular system. Following this line of thought, large efforts have been dedicated to propose new adiabatic conditions \cite{mackenzei, comparat, vertesi, tong2, tong3, jansen, tong4,du, yongye}.  
Nonetheless, recent work \cite{frasca, comparat2} indicates that the debate about the full extent of the inconsistency remains open. 

Here, by carefully examining the standard proof of QAT due to Messiah \cite{messiah}, we show in Sec. \ref{dos} that the theorem does not hold for the dual system because some key premises required in the demonstrations of the theorem are not satisfied by $H_{b}$. This confirms that only the adiabatic approximation is affected by the Marzlin-Sanders inconsistency, while the theorem is correct \cite{mackenzei}.
As a result of our analysis we show in Sec. \ref{tres} that two simple conditions can be used to determine if a system for which  Eq. (\ref{Eq1}) holds satisfies the adiabatic approximation. We illustrate our findings in Sec. \ref{cuatro} with an example. In Sec. \ref{cinco} we discuss the full extent of the inconsistency, showing that it only arises when the Hamiltonian contains resonant oscillatory terms whose amplitude goes to zero in the asymptotic limit. Finally, we give a summary and a general discussion in Sec. \ref{seis}.

\section{The quantum adiabatic theorem}\label{dos}

\subsection{Premises}\label{dosa}
The adiabatic theorem refers to the limit in which a time-dependent Hamiltonian $H_{a}(t)$ varies infinitely slowly.  The theorem gives approximate solutions to the time-dependent Schr\"odinger equation in an interval
$0\le t \le \tau$, when $\tau\rightarrow\infty$ and the total change of $H_{a}(t)$ is finite \cite{kato}. Using a scaled time variable $s$ defined by $t=\tau s$, $0 \le s \le 1$, the Schr\"odinger equation
can be written

\begin{equation}
i\hbar\frac{d \phi(s)}{ds}=\tau H_{a}(s)\phi(s) \;.
\label{schrodinger}
\end{equation}

In the following, we assume that the energy spectrum is discrete at all times as the Marzlin-Sanders paradox arises in such a context.  Hence, the following theorem 

\begin{equation}
U_{a}(s)P^{a}_{n}(0)-P^{a}_{n}(s)U_{a}(s)=O(1/\tau), \;\; \tau\rightarrow \infty, \;\forall n \;,
\label{nuevo0}
\end{equation}

\noindent
can be proven under the following additional premises \cite{messiah}:  (p1) the eigenvalues $E^{a}_{n}(s)$ are continuous functions of $s$, 
(p2) there are no eigenvalue crossings (i.e., $E^{a}_{n}(s)\ne E^{a}_{m}(s)$, $n \ne m$, $s\in [0,1]$), and (p3) the derivatives of the eigenprojectors, $dP^{a}_{n}(s)/ds$ and $d^{2}P^{a}_{n}(s)/ds^{2}$, are well defined and are piecewise continuous in the interval $0\leq s \leq 1$. For simplicity $H_{a}(s)$ is usually assumed to be independent of $\tau$ but the arguments leading to the adiabatic theorem apply similarly to the case in which $H_{a}$ depends explicitely on $\tau$, as far as premises (p1)-(p3) are uniformly fulfilled for $\tau \rightarrow \infty$ \cite{kato, nenciu}.

\subsection{Sketch of the standard proof of the theorem}\label{dosb}
Messiah's demonstration \cite{messiah} is based on the use of a virtual evolution operator, $U_{A}(s)$, defined by Kato \cite{kato},  that takes the system from a given eigenstate at the initial time to the corresponding eigenstate at future times.  A purely geometric evolution is realized by \cite{berry}

\begin{equation}
U_{A}(s)=\sum_{n}|E^{a}_{n}(s)\rangle \langle E^{a}_{n}(0)| \;,
\label{berry}
\end{equation}

\noindent
where the phases of the instantaneous eigenfunctions, $|E_{n}^{a}(s)\rangle$, are such that the parallel-transport condition $\langle E_{n}^{a}(s)|\dot{E}_{n}^{a}(s)\rangle=0$ is satisfied. Operator $U_{A}$ fulfills the so-called intertwining property

\begin{equation}
U_{A}(s)P_{n}^{a}(0)=P_{n}^{a}(s)U_{A}(s)\;,
\label{eqkato}
\end{equation}

\noindent
and therefore, QAT, Eq. (\ref{nuevo0}), can be proven by showing that

\begin{equation}
\lim_{\tau\rightarrow \infty}U_{A}^{\dagger}(s) U_{a}(s)=\Phi_{A}(s) \left[1+O(1/\tau)\right] \;,
\label{nuevo00}
\end{equation}

\noindent
where the operator $\Phi_{A}(s)$ can be expanded in terms of the eigenprojectors at $s=0$ multiplied by phase factors

\begin{equation}
\Phi_{A}(s)=\sum_{n}\exp\left[(-i\tau/\hbar) \int_{0}^{s} E_{n}^{a}(\sigma) d\sigma\right] P_{n}^{a}(0)\;.
\label{nuevo1}
\end{equation}

Messiah showed that the solution of the evolution equation for operator $\Omega_{A}\equiv U_{A}^{\dagger} U_{a}$ goes to $\Phi_{A}$ in the limit $\tau \rightarrow \infty$. For that, he made
two unitary transformation over the evolution equation in the Schr\"odinger representation. 
This procedure leads to the following Volterra integral evolution equation for operator $W_{A}\equiv\Phi_{A}^{\dagger}\Omega_{A}$

\begin{equation}
W_{A}(s)=1+(i/\hbar)\int_{0}^{s} \overline{K}_{a}(\sigma) W_{A}(\sigma)d\sigma \;,
\label{integral}
\end{equation}

\noindent
where the kernel, 

\begin{equation}
\overline{K}_{a}(s)=\Phi_{A}^{\dagger}(s) U_{A}^{\dagger}(s) K_{a}(s) U_{A}(s) \Phi_{A}(s) \;,
\label{kernel3}
\end{equation}

\noindent
depends on the following operator

\begin{equation}
K_{a}(s)=i\hbar\sum_{n}\dot{P}_{n}^{a}(s)P_{n}^{a}(s) \;,
\label{virtualh}
\end{equation}

\noindent
which is the generator of the virtual evolution given by $U_{A}$.
The completion of the proof requires showing that the integral in the Volterra equation goes to zero when $\tau\rightarrow \infty$. If that is the case, Eq. (\ref{integral}) gives $W_{A}\rightarrow 1$, and from the definition of $\Omega_{A}$ results $\Omega_{A}\rightarrow \Phi_{A}$, and finally $U_{a}\rightarrow U_{A} \Phi_{A}$.

An estimate of the integral in Eq. (\ref{integral}) can be obtained by substituting in Eq. (\ref{kernel3}) the standard expansion $K_{a}=\sum_{nm} P_{n}(0)K_{a}P_{m}(0)$, and Eq. (\ref{nuevo1}), which gives the following expression for the kernel

\begin{equation}
\overline{K}_{a}(s)=i\hbar \sum_{m\neq n}\exp\left[(i\tau/\hbar) \int_{0}^{s} \left(E_{m}^{a}(\sigma)-E_{m}^{a}(\sigma)\right) d\sigma\right]  
\langle E_{m}^{a}(s)|\dot{E}_{n}^{a}(s)\rangle |E_{m}^{a}(0)\rangle\langle E_{n}^{a}(0)|\;.
\label{bar2}
\end{equation}

\noindent
Nondiagonal elements of the kernel contain exponential factors whose phases are never stationary because they oscillate at  frequencies that increase with $\tau$, as far as the differences $E_{n}^{a}(\sigma)-E_{m}^{a}(\sigma)$ are nonzero (i.e., 
{\it if the eigenvalues are separated by gaps}). Messiah showed that the integral
in Eq. (\ref{integral}) can be expressed as a sum of two terms, which contain as a factor the operator

\begin{equation}
F_{a}(s)\equiv\int_{0}^{s} \overline{K}_{a}(\sigma) d\sigma \;.
\label{adtosketch}
\end{equation}

Thus, if $F_{a}(s)$ goes to zero, the integral in the Volterra equation will go to zero too. Due to the presence of oscillatory factors in the kernel, operator $F_{a}(s)$ approaches zero  if the elements $\langle E_{m}^{a}(s)|\dot{E}_{n}^{a}(s)\rangle$ are continuous functions of $s$, and {\it  their derivatives remain finite} for all $s$ \cite{messiah}.

\section{Identification of systems affected by the inconsistency}\label{tres}

\subsection{The dual system}\label{dosc}

Amin \cite{explanations} showed, by taking into account the adiabaticy of $S_{a}$, that the dual Hamiltonian $H_{b}$ can be expanded as

\[
H_{b}(s, \tau)=-\sum_{n}E_{n}^{a}(s) |E_{n}^{a}(0)\rangle \langle E_{n}^{a}(0)|
\]

\begin{equation}
-\frac{i \hbar}{\tau} \sum_{m\neq n}\exp\left[(-i \tau/\hbar) \int_{0}^{s} (E_{n}^{a}(\sigma)-E_{m}^{a}(\sigma)) d\sigma \right] \langle E_{m}^{a}(s)|\dot{E}_{n}^{a}(s)\rangle |E_{m}^{a}(0)\rangle \langle E_{n}^{a}(0)| \;,
\label{amineq}
\end{equation}

\noindent
where the expression given in \cite{explanations} has been rewritten here as a function of scaled time. 

Similarly, the eigenprojectors, $P_{b}$, can be written as

\[
P_{n}^{b}(s, \tau)=P_{n}^{a}(0)-\frac{i \hbar}{\tau E_{n}^{a}(s)} \sum_{m} \exp\left[(-i \tau/\hbar) \int_{0}^{s} (E_{n}^{a}(\sigma)-E_{m}^{a}(\sigma)) d\sigma \right]
\]

\begin{equation}
\times\langle E_{m}^{a}(s)|\dot{E}_{n}^{a}(s)\rangle |E_{m}^{a}(0)\rangle \langle E_{n}^{a}(0)| \;,
\label{amineqb}
\end{equation}

\noindent
which shows that the Hamiltonian and the eigenprojectors for the dual system depend explicitely on $\tau$. This fact does not invalidate {\it a priori} the adiabatic theorem as far as premises (p1)-(p3) hold when $\tau \rightarrow \infty$.  

Premises (p1) and (p2) depend only on the gaps between pairs of eigenvalues. Therefore, they are equally satisfied for $H_{a}$ and $H_{b}$ since $E_{n}^{b}(s)=-E_{n}^{a}(s)$. However, premise (p3) does not hold for $H_{b}$, since the operator-valued functions, $dP_{n}^{b}/ds$ and $d^{2}P_{n}^{b}/ds^{2}$, are not defined when $\tau \rightarrow \infty$, because the argument of the exponential in Eq. (\ref{amineqb}) goes to infinity. The explicit $\tau$-dependence of the Hamiltonian, when premise (p3) does not hold, prevents the adiabatic limit for $S_{b}$ from being reached when $\tau\rightarrow \infty$, unlike for $S_{a}$.

The adiabatic theorem holds for the dual system if the integral in the Volterra equation, Eq. (\ref{integral}), for $S_{b}$ goes to zero in the $\tau\rightarrow\infty$ limit. As explained in Sec. 
\ref{dosb} such an integral goes to zero whenever (i) the eigenvalues are separated by gaps and (ii) the elements $\langle E_{m}^{b}(s)|\dot{E}^{b}_{n}(s)\rangle$ are continuous functions of $s$ that remain finite along with their derivatives for all $s$. Thus, the adiabatic theorem would be satisfied for the dual system if (ii) holds since (i) is automatically guaranteed because $H_{b}(s)$ has the same spectral gaps as $H_{a}(s)$, which is adiabatic by hypothesis.

A wrong reasoning indicates that the integral in the Volterra equation should go to zero for system $S_{b}$ as it does for $S_{a}$, since Eqs. (\ref{berry})-(\ref{bar2}) are apparently valid for $S_{b}$ if subscript $A(a)$ is changed to $B(b)$.  Then, the exact $\Omega_{B}$ would tend to an operator analogous to Eq. (\ref{nuevo1}) and QAT would hold for $S_{b}$.
From $E_{n}^{b}(s)=-E_{n}^{a}(s)$ and $P_{n}^{a}(0)=P_{n}^{b}(0)$,  this operator is $\Phi_{B}=\Phi_{A}^{\dagger}$.

\subsection{Properties satisfied by the dual system in the $\tau\rightarrow \infty$ limit}\label{tresb}

The naivety of the previous argument is exposed by realizing that a virtual adiabatic operator for $S_{b}$ is given, in general, as $U_{B}=U_{a}^{\dagger}U_{A}\Phi $. Note that a Volterra equation analogous to Eq. (\ref{integral}) is obtained {\it only} if $U_{B}$ is a  geometric evolution operator; that is, if it can be expanded like Eq. (\ref{berry}) in terms of eigenfunctions of $H_{b}$ that satisfy 
$\langle E_{n}^{b}(s)|\dot{E}_{n}^{b}(s)\rangle=0$. The following relationship between parallel-transport eigenfunctions of $S_{a}$ and $S_{b}$, 

\begin{equation}
|E^{b}_{n}(s, \tau)\rangle=\exp \left[(-i \tau/\hbar)\int_{0}^{s} E_{n}^{a}(\sigma) d\sigma\right]U_{a}^{\dagger}(s)|E^{a}_{n}(s)\rangle \;,
\label{parallelt}
\end{equation}

\noindent
allows us to expand the virtual evolution operator, $U_{B}$, in terms of eigenfunctions of $S_{a}$, which satisfy $\langle E_{n}^{a}(s)|\dot{E}_{n}^{a}(s)\rangle=0$, as

\begin{equation}
U_{B}(s, \tau)=\sum_{n} \exp \left[(-i \tau/\hbar)\int_{0}^{s} E_{n}^{a}(\sigma) d\sigma\right]
U_{a}(s)^{\dagger}|E_{n}^{a}(s)\rangle \langle E_{n}^{a}(0)| \;.
\label{ub}
\end{equation}

\noindent
On the other hand,
the relation between the eigenprojectors of both systems

\begin{equation}
P_{n}^{b}(s, \tau)=U_{a}^{\dagger}(s)P_{n}^{a}(s)U_{a}(s) \;,
\label{relation}
\end{equation}

\noindent
gives $K_{b}(s)=U_{a}^{\dagger}(s)K_{a}(s)U_{a}(s)$, which can be substituted, along with $\Phi_{B}(s)=\Phi_{A}^{\dagger}(s)$, and Eq. (\ref{ub}),  in the analogous equation to Eq. (\ref{kernel3}) appropriate for system $S_{b}$, to obtain

\begin{equation}
\overline{K}_{b}=\Phi_{B}^{\dagger}U_{B}^{\dagger}K_{b}U_{B}\Phi_{B}=\Phi_{A}\Phi_{A}^{\dagger}U_{A}^{\dagger}U_{a}U_{a}^{\dagger}K_{a}U_{a}U_{a}^{\dagger}U_{A}\Phi_{A}\Phi_{A}^{\dagger} \;,
\label{bar}
\end{equation}

\noindent
where the $s$ dependence has been omitted for simplicity.
Finally, by using the equations, for $S_{b}$, analogous to Eqs. (\ref{berry}) and (\ref{virtualh}), $\overline{K}_b$ can be expanded as

\begin{equation}
\overline{K}_{b}(s)=i \hbar\sum_{m \neq n} \langle E_{m}^{a}(s)|\dot{E}_{n}^{a}(s)\rangle |E_{m}^{a}(0)\rangle\langle E_{n}^{a}(0)|\;,
\label{siguebar}
\end{equation}

\noindent
which does not contain oscillating functions. Hence,

\begin{equation}
F_{b}(s)\equiv\int_{0}^{s} \overline{K}_{b}(\sigma) d\sigma \nrightarrow 0 \left(\frac{1}{\tau}\right) \;,
\label{condition0}
\end{equation}

\noindent
which implies that the integral in the Volterra evolution equation does not go to zero. In other words,

\begin{equation}
\int_{0}^{s} \overline{K}_{b}(\sigma) W_{B} (\sigma) d\sigma \nrightarrow 0 \left(\frac{1}{\tau}\right) \;,
\label{condition00}
\end{equation}

\noindent
because 

\begin{equation}
\lim_{\tau \rightarrow \infty} \int_{0}^{s} \exp \left[(i\tau/\hbar) \int_{0}^{\sigma} (E_{m}^{b}(\sigma')-E_{n}^{b}(\sigma')) d\sigma' \right] \langle E_{m}^{b}(\sigma)|\dot{E}_{n}^{b}(\sigma) \rangle d\sigma =\int_{0}^{s} \langle E_{m}^{a}(\sigma)|\dot{E}_{n}^{a}(\sigma) \rangle d\sigma \nrightarrow 0 \;,
\label{condition1}
\end{equation}

\noindent
due to the cancellation of the exponential factors.

The adiabatic theorem does not hold for system $S_{b}$ because the mechanism that makes the integral in Eq. (\ref{integral}) approach zero 
for $S_{a}$ fails for $S_{b}$ in spite of the fact that both systems have the same spectral gaps. The reason is that  
the oscillations of the elements $\langle E_{m}^{b}(s)|\dot{E}_{n}^{b}(s)\rangle$ cancel the terms that arise from the gap condition. We emphasize that this does not imply an inconsistency in QAT  since system $H_{b}$ does not satisfy the hypotheses of the theorem. 
Here, we prove that systems for which the effects of the gap condition are canceled can be easily identified by simple conditions. Also, we will show that such conditions can be applied without modification to identify systems that do not satisfy the adiabatic approximation in spite of satisfying Eq.~(\ref{Eq1}).

By substituting the relationship between the eigenprojectors of systems $S_{a}$ and $S_{b}$, Eq. (\ref{relation}), into the intertwining property,  Eq. (\ref{eqkato}), which $U_{A}$ satisfies due to the adiabaticity of $S_{a}$, we get

\begin{equation}
\left[U_{a}^{\dagger}(s)U_{A}(s)\right] P^{b}_{n}(0)\left[U_{A}(s)^{\dagger}U_{a}(s)\right]=P^{b}_{n}(s, \tau) \;.
\label{final3}
\end{equation}

\noindent
By taking the $\tau\rightarrow \infty$ limit, and substituting Eq. (\ref{nuevo00}) into the previous expression, we obtain

\begin{equation}
\lim_{\tau \rightarrow \infty}P_{n}^{b}(s, \tau)=P_{n}^{b}(0)[1+O(1/\tau)]^{2},  \; \forall n,\;s\in[0,1]\;,
\label{adiacon2}
\end{equation}

\noindent
which holds for systems affected by the Marzlin-Sanders inconsistency; that is, for systems that satisfiy Eq. (\ref{Eq1}) but do not evolve adiabatically. However, Eq. (\ref{adiacon2}), is 
ambiguous and therefore is not a sufficient condition. In effect, a system $S_{c}$ with Hamiltonian $-H_{b}$ has the same eigenprojectors but, being the Heisenberg representation of $H_{a}$, is trivially adiabatic. This ambiguity can be eliminated by finding some other property that holds for $S_{b}$ but not for $S_{c}$. The kernel for $H_{c}=-H_{b}$ is

\begin{equation}
 \overline{K}_{c}(s)=\Phi_{A}^{\dagger}\Phi_{A}^{\dagger}U_{A}^{\dagger}K_{a}U_{A}\Phi_{A}\Phi_{A}\;,
\label{last}
\end{equation}

\noindent
which shows that the phase cancellation that produces the inconsistency for system $S_{b}$ does not occur for system $S_{c}$.  The kernel expansion for $S_{c}$ contains oscillatory factors that are  related to those of $S_{a}$ by

\[
\exp \left[(i\tau/\hbar) \int_{0}^{\sigma} (E_{m}^{c}(\sigma')-E_{n}^{c}(\sigma')) d\sigma' \right]
\langle E_{m}^{c}(\sigma)|\dot{E}_{n}^{c}(\sigma) \rangle
\]

\begin{equation}
=\exp \left[(2i\tau/\hbar) \int_{0}^{\sigma} (E_{m}^{a}(\sigma')-E_{n}^{a}(\sigma')) d\sigma' \right]
\langle E_{m}^{a}(\sigma)|\dot{E}_{n}^{a}(\sigma) \rangle \;,
\label{conditionextra2}
\end{equation}

\noindent
where the eigenvectors $|E_{n}^{a}\rangle$, $|E_{n}^{c}\rangle$, fulfill the parallel-transport condition.  Since system $S_{a}$ is adiabatic by hypothesis, the integral of the expression at the right-hand side of Eq. (\ref{conditionextra2})  goes to zero as $1/\tau$, which implies, for $\tau\rightarrow \infty$,

\begin{equation}
\int_{0}^{s} \exp \left[(i\tau/\hbar) \int_{0}^{\sigma} (E_{m}^{c}(\sigma')-E_{n}^{c}(\sigma')) d\sigma' \right]
\langle E_{m}^{c}(\sigma)|\dot{E}_{n}^{c}(\sigma) \rangle d\sigma=O(1/\tau) \;,
\label{condition2}
\end{equation}

\noindent
while, as shown in Eq. (\ref{condition1}),  the analogous integral for system $S_{b}$ does not go to zero. 

In conclusion, if the eigenprojectors of a given Hamiltonian fulfill Eq.(\ref{adiacon2}), the traditional adiabatic conditions, Eq. (\ref{Eq1}), are insufficient to determine the adiabaticity of the time evolution. In such a case, only if the integral in the left-hand side of Eq. (\ref{condition1}) goes to zero, will the evolution be adiabatic.

\subsection{Properties satisfied by the dual system for finite $\tau$}\label{tresa}

The adiabatic approximation, as mentioned in the Introduction, is the set of conditions under which the time evolution of a system that evolves during a finite time interval, is still approximately adiabatic. A system for which the adiabatic theorem holds when $\tau \rightarrow \infty$ could fail to evolve adiabatically for a certain set of finite values of the parameter $\tau$. However, if the adiabatic theorem does not hold when $\tau \rightarrow \infty$, the evolution cannot be approximately adiabatic for any finite $\tau$. 
Since the dual system does not obey the adiabatic theorem it does not obey the adiabatic approximation either. However, the failure to obey premise (p3), which invalidates the theorem, refers to the strict adiabatic limit and cannot be used to determine whether the adiabatic approximation does not hold. Next, we show that there exist common conditions that apply to both the adiabatic theorem and the adiabatic approximation, from which we will obtain properties that can be used to determine the validity of Eq. (\ref{Eq1}).

Previous studies of the Marzlin-Sanders inconsistency have considered mainly the development of new criteria of validity for the adiabatic approximation \cite{mackenzei, comparat, tong2, tong3, jansen, vertesi}. In the following, we use directly the variable $t$ for time instead of $s$, as it is traditional in the analysis of the adiabatic approximation. Appropriate criteria must identify unambiguously if a Hamiltonian that acts during a finite time interval varies slowly enough for the state vector at time $t$ to be approximated by the eigenstate $|E_{n}(t)\rangle$ except for a phase factor, if the state vector at initial time was $|E_{n}(0)\rangle$.  

In the $\tau \rightarrow \infty$ limit, conditions, Eqs. (\ref{condition1}), and (\ref{adiacon2}), are sufficient to indicate if a given Hamiltonian, $H_{b}$,  does not satisfy the adiabatic theorem.  These same conditions can identify systems that satisfy Eq. (\ref{Eq1}) but that, however, do not satisfy the adiabatic approximation. In effect, if $H_{a}$ satisfies the adiabatic approximation, we get

\begin{equation}
U_{a}(t)\approx U_{A}(t)\Phi_{A}(t) \;.
\label{adiapro0}
\end{equation}

\noindent
Substituting this expression in Eq. (\ref{final3}), which approximately holds for finite but long enough $\tau$, we obtain

\begin{equation}
P_{n}^{b}(t)\approx P_{n}^{b}(0) \;, \; \forall t \;.
\label{adiapro1}
\end{equation}

On the other hand, Eq. (\ref{condition1})  holds too, because the matrix elements 
$\langle E_{m}^{b}(t)|\dot{E}_{n}^{b}(t)\rangle$ cancel the exponentials arising from the gap condition for finite $\tau$ as in the $\tau \rightarrow \infty$ case. Finally, if the adiabatic approximation is obeyed, the integral in Eq. (\ref{condition2})  will be small.

\section{Example}\label{cuatro}

The previous results can be illustrated by analyzing a model that has been extensively studied in relation to the Marzlin-Sanders inconsistency. This model consists of a spin-half particle in a rotating magnetic field whose Hamiltonian is \cite{tong}

\begin{equation}
H_{a}(t)=-\omega_{0}/2(\sigma_{x}\sin\theta\cos \omega t+\sigma_{y}\sin \theta \sin \omega t+\sigma_{z} \cos\theta) \;,
\label{hamiltonian}
\end{equation}

\noindent
where $\omega$ and $\omega_{0}$ are constants, while the $\sigma_{i}'s$ are Pauli matrices. The instantaneous eigenvalues are $E_{1}^{a}(t)=\omega_{0}/2$ and $E_{2}^{a}(t)=-\omega_{0}/2$, while the instantaneous eigenvectors are \cite{tong}

\begin{equation}
|\overline{E}_{1}^{a}(t)\rangle= \left( \begin{array}{c}
e^{-i \omega t/2} \sin \frac{\theta}{2} \\
-e^{i \omega t/2} \cos \frac{\theta}{2} \end{array} \right) \;,
\label{eeigen1}
\end{equation}

\noindent
and

\begin{equation}
|\overline{E}_{2}^{a}(t)\rangle= \left( \begin{array}{c}
e^{-i \omega t/2} \cos \frac{\theta}{2} \\
e^{i \omega t/2} \sin \frac{\theta}{2} \end{array} \right) \;,
\label{eeigen2}
\end{equation}

\noindent
which do not obey the parallel-transport phase condition.

The Hamiltonian, Eq. (\ref{hamiltonian}) can be written as a function of scaled time $s=2\pi t/\tau=\omega t$

\begin{equation}
H_{a}(s)=-\omega_{0}/2(\sigma_{x}\sin\theta\cos s+\sigma_{y}\sin \theta \sin s+\sigma_{z} \cos\theta) \;.
\label{hamilredu}
\end{equation}

\noindent
Eigenfunctions of $H_{a}(s)$ that obey the parallel-transport condition are given, after transforming Eqs. (\ref{eeigen1}) and (\ref{eeigen2}) to scaled time, by 
$|E_{n}\rangle=e^{-\int_{0}^{s}\langle \overline{E}_{n}|\dot{\overline{E}}_{n}\rangle d\sigma} |\overline{E}_{n}\rangle$. Thus, we get

\begin{equation}
|E_{1}^{a}(s)\rangle= e^{-i (s/2) \cos\theta}\left( \begin{array}{c}
e^{-i s/2} \sin \frac{\theta}{2} \\
-e^{i s/2} \cos \frac{\theta}{2} \end{array} \right) \;,
\label{eeeigen1}
\end{equation}

\noindent
and

\begin{equation}
|E_{2}^{a}(s)\rangle= e^{i (s/2) \cos \theta}\left( \begin{array}{c}
e^{-i s/2} \cos \frac{\theta}{2} \\
e^{i s/2} \sin \frac{\theta}{2} \end{array} \right) \;.
\label{eeeigen2}
\end{equation}

Finally, note that the adiabatic limit for $H_{a}(s)$ corresponds to $\tau\rightarrow \infty$, which is equivalent to $\omega \rightarrow 0$. Such a limit is not physically realizable, but that is not relevant for the illustration of the inconsistency.

\subsection{The  $\omega \rightarrow 0$ limit}\label{cuatroa}

We showed above that a dual Hamiltonian 
$H_{b}(s, \tau)=-U_{a}^{\dagger}(s, \tau) H_{a}(s) U_{a}(s, \tau)$ does not satisfy, in general, the adiabatic theorem. Let us study the case corresponding to $H_{a}$ given by Eq. (\ref{hamilredu}).  Hamiltonian $H_{b}$ can be written as

\begin{equation}
H_{b}(s, \tau)=-\sum_{n} E_{n}^{a} P_{n}^{b}(s, \tau ) \;,
\label{expansioneigenp}
\end{equation}

\noindent
where the instantaneous eigenprojectors are given by $P_{n}^{b}=U_{a}^{\dagger}P_{n}^{a} U_{a}$.  Transforming the expression for $U_{a}$, given by Tong \cite{tong}, to scaled time, we get

\begin{equation}
U_{a}(s, \tau) = \left( \begin{array}{cc}
\left(\cos \frac{\overline{\omega}s}{2\omega}+ i \frac{\omega+\omega_{0} \cos \theta}{\overline{\omega}}\sin \frac{\overline{\omega} s}{2\omega}\right) e^{-i s/2} &
i \frac{\omega_{0} \sin\theta}{\overline{\omega}} \sin \frac{\overline{\omega}s}{2 \omega} e^{-i s/2} \\
i \frac{\omega_{0} \sin\theta}{\overline{\omega}} \sin \frac{\overline{\omega}s}{2 \omega}  e^{i s/2} &
\left(\cos  \frac{\overline{\omega}s}{2\omega}- i \frac{\omega+\omega_{0} \cos \theta}{\overline{\omega}} \sin \frac{\overline{\omega}s}{2\omega}\right) e^{i s/2} \end{array} \right) \;,
\label{matrizu}
\end{equation}

\noindent
where $\overline{w}=\sqrt{w_{0}^{2}+w^{2}+2 w w_{0} \cos \theta}$. On the other hand, the following matrix representation for $P_{1}^{a}$ can be constructed from the instantaneous eigenvector, Eqs. (\ref{eeeigen1}),  

\begin{equation}
P_{1}^{a}(s)= \left( \begin{array}{cc}
\sin^{2} \frac{\theta}{2}  & -\sin \frac{\theta}{2} \cos\frac{\theta}{2} e^{-is} \\
-\sin \frac{\theta}{2} \cos\frac{\theta}{2} e^{is} & \cos^{2} \frac{\theta}{2}  \end{array} \right) \;. 
\label{matrixp}
\end{equation}

\noindent
The matrix representation for the instantaneous eigenprojector, $P_{1}^{b}(s)$, can be calculated from the previous two equations and has a complicated form. For example, one of its matrix elements is

\[
[P_{1}^{b}(s, \tau)]_{12} = \sin^{2} \left(\frac{\overline{\omega}}{2\omega} s\right) \left[\frac{\omega_{0}\left(\omega+\omega_{0}\cos\theta\right)}{\overline{\omega}^{2}} \sin \theta \sin^{2}
\frac{\theta}{2}-\left(\frac{\omega_{0}\sin \theta}{\overline{\omega}}\right)^{2}  \sin\frac{\theta}{2} \cos\frac{\theta}{2} \right.
\]

\[
\left.+\left( \frac{\omega+\omega_{0}\cos\theta}{\overline{\omega}}\right)^{2} \sin\frac{\theta}{2}\cos\frac{\theta}{2}-\frac{\omega_{0}\sin \theta \left(\omega+\omega_{0} \cos \theta\right)}{\overline{\omega}^{2}} \cos^{2} \frac{\theta}{2}\right]
-\cos^{2} \left(\frac{\overline{\omega}}{2\omega} s\right) \sin \frac{\theta}{2}\cos \frac{\theta}{2}
\]

\begin{equation}
+i \sin \left(\frac{\overline{\omega}}{2\omega} s\right) \cos \left(\frac{\overline{\omega}}{2\omega} s\right) \left[\frac{\omega_{0}}{\overline{\omega}} \sin\theta \sin^{2} \frac{\theta}{2}+
2 \frac{\omega+\omega_{0}\cos\theta}{\overline{\omega}} \sin\frac{\theta}{2} \cos \frac{\theta}{2} -\frac{\omega_{0}}{\overline{\omega}} \sin \theta \cos^{2} \frac{\theta}{2} \right] \;.
\label{proele}
\end{equation}

We showed in Eq. (\ref{amineqb}) that the eigenprojectors of the dual system contain terms that, although they oscillate infinitely fast, go to zero in the 
$\tau\rightarrow \infty$ limit.  Apparently, this is not the case for $[P_{1}^{b}]_{12}$. However, Eq. (\ref{proele}) can be written, by using well known trigonometric identities such as

\[
[P_{1}^{b}(s, \tau)]_{12} =\sin^{2}\left(\frac{\overline{\omega}}{2\omega} s\right) \left[-\left(\frac{\omega_{0}}{\overline{\omega}}\right)^{2} \sin\frac{\theta}{2}\cos\frac{\theta}{2}+
\left(\frac{\omega}{\overline{\omega}}\right)^{2} \sin\frac{\theta}{2}\cos\frac{\theta}{2}\right]
\]

\begin{equation}
-\cos^{2} \left(\frac{\overline{\omega}}{2\omega} s\right) \sin \frac{\theta}{2}\cos \frac{\theta}{2}
+2 i \sin \left(\frac{\overline{\omega}}{2\omega} s\right) \cos \left(\frac{\overline{\omega}}{2\omega} s\right) \frac{\omega}{\overline{\omega}}\sin\frac{\theta}{2}\cos\frac{\theta}{2} \;.
\label{proele2}
\end{equation}

\noindent
By taking into account that $\omega\rightarrow 0$ implies $\overline{\omega}\rightarrow \omega_{0}$, we get

\[
\lim_{\omega\rightarrow 0}[P_{1}^{b}(s,\tau)]_{12}=-\sin\frac{\theta}{2} \cos\frac{\theta}{2} \left[1-\sin^{2}\left(\frac{\overline{\omega}}{2\omega} s\right) \left(\frac{\omega}{\overline{\omega}}\right)^{2}+2 i \sin \left(\frac{\overline{\omega}}{2\omega} s\right) \cos \left(\frac{\overline{\omega}}{2\omega} s\right) \frac{\omega}{\overline{\omega}}\right]
\]

\begin{equation}
=[P_{1}^{b}(0)]_{12}\left[1+O(\omega)+O(\omega^{2})\right] \;.
\label{proelemfinal}
\end{equation}

\noindent
Following the same procedure for the other  elements of the eigenprojector matrix we obtain

\begin{equation}
\lim_{\omega \rightarrow 0}  P_{1}^{b} (s, \tau)  = \left( \begin{array}{cc}
\sin^{2} \frac{\theta}{2} & -\sin \frac{\theta}{2} \cos\frac{\theta}{2} \\
-\sin \frac{\theta}{2} \cos \frac{\theta}{2} & \cos^{2} \frac{\theta}{2} \end{array} \right) \left[1+O\left(\omega\right)  {\rm M}\left(\frac{\overline{\omega}}{2\omega} s\right)
+O(\omega^{2}) {\rm N}\left(\frac{\overline{\omega}}{2\omega}s\right)\right]\;,
\label{proofeq1}
\end{equation}

\noindent
where {\rm M} and {\rm N} represent matrices whose elements are functions that oscillate infinitely fast. However, they are multiplied by terms that go to zero in the $\omega \rightarrow 0$ limit.
Equation (\ref{proofeq1}) shows that $P_{1}^{b}(s)$ [and $P_{2}^{b}(s)$, which is not given here], in the $\omega \rightarrow 0$ limit, tend to $P_{1}^{b}(0)$ [and $P_{2}^{b}(0)$]. Note also that the instantaneous eigenprojectors do not contain several independent time scales.  The time dependence has the form $\overline{\omega}s/(2\omega)$, which suggests the use of a new scaled variable $s'=\overline{\omega} s/(2 \omega)$. Also, the eigenvalues are constant, for this particular example, and consequently the Hamiltonian can be written as a function of $s'$ too. This shows that the adiabatic limit for $H_{b}$, for which QAT would hold, requires $\overline{\omega}/(2 \omega)\rightarrow 0$ in addition to $\omega \rightarrow 0$. This limit is physically meaningless.

As explained above, Hamiltonian $H_{c}=-H_{b}$, has the same eigenprojectors as $H_{b}$ but, is adiabatic. Both systems can be distinguished by evaluating 

\begin{equation}
\int_{0}^{s} \exp \left[\left(i/ \omega\right) \int_{0}^{\sigma} (E_{1}(\sigma')-E_{2}(\sigma')) d\sigma' \right] \langle E_{1}(\sigma)|\dot{E}_{2}(\sigma) \rangle d\sigma \;,
\label{resolutivo}
\end{equation}

\noindent
where $E_{n}(s)=-E_{n}^{a}(s)$ for $H_{b}$ and $E_{n}(s)=E_{n}^{a}(s)$ for $H_{c}$. From the relationship between the eigenfunctions of both systems and those of $H_{a}$ we have

\begin{equation}
\langle E_{1}(\sigma)|\dot{E}_{2}(\sigma)\rangle=\exp\left[(i/\omega) \int_{0}^{\sigma} (E_{1}^{a}(\sigma')-E_{1}^{a}(\sigma'))d\sigma'\right]\langle E_{1}^{a}(\sigma)|\dot{E}_{2}^{a}(\sigma)\rangle \;,
\label{resolutivo2}
\end{equation}

\noindent
and taking into account 
 
\begin{equation}
\langle E_{1}^{a}(s)|\dot{E}_{2}^{a}(s)\rangle =-\frac{i}{2} \sin\theta e^{i s\cos\theta} \;,
\label{schwinger1}
\end{equation}

\noindent
Eq. (\ref{resolutivo}) becomes for $H_{b}$

\[
\int_{0}^{s} \exp \left[\left(i/ \omega\right) \int_{0}^{\sigma} (E_{1}^{b}(\sigma')-E_{2}^{b}(\sigma')) d\sigma' \right] \langle E_{1}^{b}(\sigma)|\dot{E}_{2}^{b}(\sigma) \rangle d\sigma
\]

\begin{equation}
=\int_{0}^{s}\langle E_{1}^{a}(\sigma)|\dot{E}_{2}^{a}(\sigma) \rangle d\sigma =\frac{1}{2} (1-e^{i s \cos \theta}) \tan \theta \;,
\label{result2}
\end{equation}

\noindent
which shows that $S_{b}$ is not adiabatic unless $\theta=0, \mod \pi$. However, for  $H_{c}$, we have

\[
\int_{0}^{s} \exp \left[\left(i/\omega\right) \int_{0}^{\sigma} (E_{1}^{c}(\sigma')-E_{2}^{c}(\sigma')) d\sigma' \right] \langle E_{1}^{c}(\sigma)|\dot{E}_{2}^{c}(\sigma) \rangle d \sigma
\]

\[
=\int_{0}^{s} \exp \left[\left(2 i/ \omega\right) \int_{0}^{\sigma} (E_{1}^{a}(\sigma')-E_{2}^{a}(\sigma')) d\sigma' \right] \langle E_{1}^{a}(\sigma)|\dot{E}_{2}^{a}(\sigma) \rangle d\sigma
\]

\begin{equation}
=-\frac{i}{2}\sin \theta   \int_{0}^{s} e^{i \left(\frac{2\omega_{0}}{\omega}+\cos\theta \right) \sigma}  d\sigma
=\frac{\omega \sin\theta \left[1-\exp\left[ \frac{i s \left(2\omega_{0}+\omega \cos \theta\right)}{\omega}\right]\right]}{4 \left(\omega_{0}+2\omega \cos\theta\right)} \;,
\label{result}
\end{equation}

\noindent
which goes to zero in the limit $\omega \rightarrow 0$,  indicating that the adiabatic theorem holds for system $S_{c}$, like for system $S_{a}$, regardless of the values of the parameters $\omega_{0}$ and $\theta$. This example shows the validity of Eqs. (\ref{condition1}), and (\ref{adiacon2}) to identify if a Hamiltonian $H_{b}$ is
related to an adiabatic Hamiltonian $H_{a}$ through $H_{b}=-U_{a}^{\dagger}H_{a}U_{a}$.

\subsection{The finite $\omega$ case}\label{cuatrob}

System $S_{a}$ satisfies the adiabatic approximation if $\omega$ is small compared to $\omega_{0}$, which implies $\overline{\omega}\approx\omega_{0}$. If we employ this approximation in the expression for the matrix element of the eigenprojector $P_{1}^{b}(t)$, which can be obtained writing Eq. (\ref{proele}) as a function of $t$, we get

\begin{equation}
[P_{1}^{b}(t)]_{12}\approx -\sin \frac{\theta}{2} \cos\frac{\theta}{2}\left[\sin^{2}\left(\frac{\omega_{0}}{2}t\right)+ \cos^{2} \left( \frac{\omega_{0}}{2} t \right)  \right]+i\sin\left(\frac{\omega_{0}}{2}t\right) \cos \left( \frac{\omega_{0}}{2} t \right) \times 0 =-\sin\frac{\theta}{2} \cos \frac{\theta}{2} \;.
\label{check1}
\end{equation}

\noindent
The other matrix elements can be simplified in the same way, so we finally obtain $P_{1}^{b}(t)\approx P_{1}^{b}(0)$, which proves that Eq. (\ref{adiapro1}) holds for this system.

On the other hand, for finite $\omega$, i.e., for finite $\tau$, Eq. (\ref{result2})  becomes

\begin{equation}
\int_{0}^{t}\langle E_{1}^{a}(\sigma)|\dot{E}_{2}^{a}(\sigma) \rangle d\sigma =\frac{1}{2} (1-e^{i \omega t \cos \theta} ) \tan \theta \;,
\label{result2apro}
\end{equation}

\noindent
which indicates that system $S_{b}$ satisfies the adiabatic approximation only for $\theta=0$, mod $\pi$,
while the integral in Eq. (\ref{result}) is 

\[
\int_{0}^{t} \exp \left[2 i\int_{0}^{\sigma} (E_{1}^{a}(\sigma')-E_{2}^{a}(\sigma')) d\sigma' \right] \langle E_{1}^{a}(\sigma)|\dot{E}_{2}^{a}(\sigma) \rangle d\sigma
\]

\begin{equation}
=\frac{\omega \sin\theta \left[1-\exp\left( i t \left(2\omega_{0}+\omega \cos \theta\right)\right)\right]}{4\left(\omega_{0}+2\omega \cos\theta\right)} \;,
\label{ultima}
\end{equation}

\noindent
which is small for $\omega_{0} >> \omega$; that is, system $S_{c}$ satisfies the adiabatic approximation under the same conditions as system $S_{a}$, while $S_{b}$ does not.

\section{Full extent of the Marzlin-Sanders inconsistency}\label{cinco}

Traditional adiabatic conditions, given by Eq. (\ref{Eq1}), hold for the dual Hamiltonian defined by Tong \cite{tong}. However, the time evolution driven by $H_{b}$ is not adiabatic. We have proven that the ultimate reason behind this inconsistency is the existence, in the Hamiltonian, of resonant terms that go asymptotically to zero. This makes the gap condition, usually invoked to guarantee adiabatic behavior in the $\tau \rightarrow \infty$ limit,  become irrelevant. These peculiarities of Tong's dual Hamiltonian, $H_{b}$, are due to its special relationship with an adiabatic Hamiltonian, $H_{a}$. In effect, both Hamiltonians are connected by a unitary transformation and a sign change. The transforming unitary operator is the exact time-evolution operator, $U_{a}$, for the adiabatic Hamiltonian. It seems pertinent to investigate the status of the inconsistency for Hamiltonians, $H_{x}$, that are related to an adiabatic Hamiltonian, $\widetilde{H}_{a}$, through a more general unitary transformation, with or without a sign change (i.e., $H_{x}=\pm U_{x}^{\dagger}\widetilde{H}_{a} U_{x}$, where $U_{x}\neq U_{a}$).

\subsection{The family $H_{x}$ is generic}\label{cincob}

At first sight it may seem that $H_{x}$ is affected by a fundamental restriction, since its eigenvalues must satisfy $E_{n}^{x}=\pm E_{n}^{a}$. However, we argue here that all possible cases of interest can be studied within this approach. Specifically, $H_{x}$ can contain generic oscillatory terms, as can be seen by turning upside down the previous argument relating $H_{x}$ and $\widetilde{H}_{a}$. In other words, instead of arguing that generic oscillatory Hamiltonians can be obtained by unitary transformation of a particular adiabatic Hamiltonian, it can be argued that for a given Hamiltonian, $H_{x}$, that by hypothesis is generic, appropriate unitary transformations lead to an adiabatic Hamiltonian. These unitary transformation are nonperturbative (i.e., they are not close to the identity). The resulting adiabatic Hamiltonian can be understood as an effective Hamiltonian that contains the effect of resonances \cite{kamfloquet}.

Therefore, given a  Hamiltonian $H_{x}(s, \tau)$ that is generic, in the sense that it may contain strong oscillatory terms (instead of the weak resonances that appear in the dual Hamiltonian used in the Marzlin-Sanders inconsistency), there exists an unitary operator, $U_{x}(s, \tau)$, that depends explicitely on $\tau$, such that the transformed Hamiltonian, $\pm U_{x}(s,\tau)H_{x}(s,\tau)U_{x}^{\dagger}(s,\tau)=\widetilde{H}_{a}(s)$, is adiabatic. Note that $\widetilde{H}_{a}$ is not necessarily the same adiabatic Hamiltonian, $H_{a}$, studied in previous sections, since $H_{x}\neq H_{b}$.  Note also that the unitary operator $U_{x}(s, \tau)$ {\it is not} the evolution operator for systems $\widetilde{S}_{a}$. Thus, this procedure does not impose {\it a priory} any restriction on $H_{x}$. However, 
the unitary operator, $U_{x}$, is somehow restricted, because it must transform the generic $H_{x}$ into an adiabatic Hamiltonian. This transformation can be very complicated but it is always possible for Hermitian matrices because they can be diagonalized. Thus, for a given $H_{x}$, there is always a similarity transformation with an unitary operator, $U_{x}=U_{D}$, that converts $H_{x}$ into a diagonal matrix, $D$. The time evolution driven by $H_{x}$, for an initial function $\Phi$, can be expressed, in terms of $U_{D}$, as \cite{tannor}

\begin{equation}
i\hbar \frac{\partial}{\partial s} \Psi(s)=\tau D(s)\Psi(s)-i\hbar U_{D}^{\dagger}(s, \tau)\frac{\partial U_{D}(s,\tau)}{\partial s} \Psi(s) \;,
\label{tannor2}
\end{equation}

\noindent
where $\Psi=U_{D}^{\dagger} \Phi$. When $H_{x}$ contains strong oscillatory terms the derivative of $U_{D}$ will be large and the time evolution is not adiabatic. However, the evolution driven by Hamiltonian $D$ is

\begin{equation}
i\hbar \frac{\partial}{\partial s} \Psi(s)=\tau D(s)\Psi(s) \;,
\label{tannor3}
\end{equation}

\noindent
which is adiabatic since $D$ does not contain off-diagonal couplings. This proves that a generic $H_{x}$ can be related by an unitary transformation, driven by $U_{x}$, to an adiabatic Hamiltonian $\widetilde{H}_{a}$. Strictly, the proof implies that, at least for $U_{x}=U_{D}$, and $\widetilde{H}_{a}=D$, the transformation exists. However, it is clear that other adiabatic Hamiltonians can be generated from the same $H_{x}$ by using less restrictive $U_{x}$ operators.

\subsection{General relations between $H_{x}$ and $\widetilde{H}_{a}$}\label{cincoc}

A generic unitary operator $U_{x}$ is {\it always} the evolution operator for an unknown Hamiltonian $\widetilde{H}\neq \widetilde{H}_{a}$

\begin{equation}
i\hbar\frac{dU_{x}(s,\tau)}{ds}=\tau\widetilde{H}(s,\tau) U_{x}(s,\tau) \;.
\label{generalU}
\end{equation}

\noindent
Parallel-transport eigenfunctions, $|E_{n}^{x}\rangle$, for $H_{x}$, can be obtained from eigenfunctions with arbitrary phase, $|\overline{E}_{n}^{x}\rangle$, as follows

\begin{equation}
|E_{n}^{x}\rangle=\exp\left[-\int_{0}^{s} \langle \overline{E}_{n}^{x}|\dot{\overline{E}}^{x}_{n}\rangle d\sigma\right] |\overline{E}_{n}^{x}\rangle \;.
\label{paralleltrans}
\end{equation}

\noindent
Since we can chose $|\overline{E}_{n}^{x}\rangle=U_{x}^{\dagger}|\widetilde{E}_{n}^{a}\rangle$,  we get, by using Eq. (\ref{generalU}),

\begin{equation}
|E_{n}^{x}(s,\tau)\rangle =\exp\left[(-i\tau/\hbar) \int_{0}^{s} \langle \widetilde{E}_{n}^{a}(\sigma)|\widetilde{H}(\sigma,\tau)|\widetilde{E}_{n}^{a}(\sigma)\rangle d\sigma \right] 
U_{x}^{\dagger}(s,\tau)  |\widetilde{E}_{n}^{a}(s)\rangle \;.
\label{parallelx}
\end{equation}

\noindent
Thus, for $n\ne m$,

\[
\langle E_{m}^{x}(s,\tau)|\dot{E}_{n}^{x}(s,\tau)\rangle=\left((i\tau/\hbar)\langle \widetilde{E}_{m}^{a}(s)|\widetilde{H}(s,\tau)|\widetilde{E}_{n}^{a}(s)\rangle+\langle \widetilde{E}_{m}^{a}(s)|\dot{\widetilde{E}_{n}^{a}}(s)\rangle\right)
\]

\begin{equation}
\times \exp\left[(i\tau/\hbar) \int_{0}^{s} \left(\langle \widetilde{E}_{m}^{a}(\sigma)|\widetilde{H}(\sigma,\tau)|\widetilde{E}_{m}^{a}(\sigma)\rangle 
-\langle \widetilde{E}_{n}^{a}(\sigma)|\widetilde{H}(\sigma,\tau)|\widetilde{E}_{n}^{a}(\sigma)\rangle\right) d\sigma \right]  \;.
\label{sigueparallelx}
\end{equation}

The relation between geometric evolution operators for $S_{x}$ and $\widetilde{S}_{a}$ contains, in general, an additional phase factor, $\Phi$, i.e., $U_{X}=U_{x}^{\dagger}\widetilde{U}_{A}\Phi$. Thus, the kernel, Eq. (\ref{kernel3}), for $S_{x}$ is

\begin{equation}
\overline{K}_{x}=\Phi_{X}^{\dagger}\Phi^{\dagger} \widetilde{U}_{A}^{\dagger}U_{x}K_{x} U_{x}^{\dagger} \widetilde{U}_{A}\Phi \Phi_{X} \;,
\label{genkernels}
\end{equation}

\noindent
where  $\Phi_{X}=\widetilde{\Phi}_{A}^{\dagger}$ if $\widetilde{H}_{a}=-U_{x}H_{x}U_{x}^{\dagger}$ or $\Phi_{X}=\widetilde{\Phi}_{A}$ if $\widetilde{H}_{a}=U_{x}H_{x}U_{x}^{\dagger}$. 
Operator $\Phi_{X}$ gets canceled, in the kernel, if $\Phi=\widetilde{\Phi}_{A}$ or $\Phi=\widetilde{\Phi}_{A}^{\dagger}$ depending on the sign of the transformation connecting $H_{x}$ and $\widetilde{H}_{a}$. In both cases, the resulting $U_{X}$ is not a geometric operator, except in special cases, because the parallel-transport phase fixing condition is not satisfied for eigenfunctions 
$|E_{n}^{x}\rangle=U_{x}^{\dagger}\widetilde{\Phi}_{A}|\widetilde{E}_{n}^{a}\rangle$ or 
$|E_{n}^{x}\rangle=U_{x}^{\dagger}\widetilde{\Phi}_{A}^{\dagger}|\widetilde{E}_{n}^{a}\rangle$. Hence, the kernel for a generic $S_{x}$
contains oscillatory terms. Also, Eq. (\ref{genkernels}) depends on the sign of the transformation that relates $H_{x}$ to $\widetilde{H}_{a}$:

\[
\overline{K}_{x}(s, \tau)=-\tau\sum_{m\ne n} \langle \widetilde{E}_{m}^{a}(s)|\widetilde{H}(s,\tau)+(i\hbar/\tau)(d/ds)|\widetilde{E}_{n}^{a}(s)\rangle |\widetilde{E}_{m}^{a}(0)\rangle \langle \widetilde{E}_{n}^{a}(0)|
\]

\begin{equation}
\times  \exp\left[\left(i\tau/\hbar\right) \int_{0}^{s} \left(\langle \widetilde{E}_{m}^{a}(\sigma)|\widetilde{H}(\sigma,\tau)\pm \widetilde{H}_{a}(\sigma)|\widetilde{E}_{m}^{a}(\sigma)\rangle-\langle \widetilde{E}_{n}^{a}(\sigma)|\widetilde{H}(\sigma,\tau) \pm \widetilde{H}_{a}(\sigma)|\widetilde{E}_{n}^{a}(\sigma)\rangle\right) d\sigma \right] \;.
\label{big}
\end{equation}

We will use these relations to study the adiabaticity of $S_{x}$ based on the characteristics of $U_{x}$. Then, we will compare the results with the left-hand side of Eq. (\ref{Eq1}) for $S_{x}$. For that, it is necessary to write Eq. (\ref{sigueparallelx}) as a function of usual time $t$ instead of $s$, which gives

\begin{equation}
\left|\frac{\langle E_{m}^{x}(t)|\dot{E}_{n}^{x}(t)\rangle}{E_{n}^{x}(t)-E_{m}^{x}(t)}\right|=\left|\frac{(i/\hbar)\langle \widetilde{E}_{m}^{a}(t)|\widetilde{H}(t)|\widetilde{E}_{n}^{a}(t)\rangle+
\langle \widetilde{E}_{m}^{a}(t)|\dot{\widetilde{E}}_{n}^{a}(t)\rangle}{\widetilde{E}_{n}^{a}(t)-\widetilde{E}_{m}^{a}(t)}\right| \;.
\label{Eq1x}
\end{equation}

We present, in the following, heuristic arguments instead of rigurous theorems. Thus, this analysis should be seen only as a sketch for future developments of the present work, whose core is the material contained in Sec. \ref{tres}.

\subsection{Status of the inconsistency for various cases}\label{cincod}

\subsubsection{$H_{x}$ does not contain oscillatory terms}\label{cincod1}

As $H_{x}$ has, by hypothesis, spectral gaps, if it does not contain oscillatory terms, premises (p1)-(p3) in Sec. \ref{dos} hold. Then, in the $\tau \rightarrow \infty$ limit, the evolution is adiabatic . Comparat \cite{comparat} showed that in this case Eq. (\ref{Eq1}) holds. Thus, no inconsistency {\it a la} Marlinz-Sanders takes place. In effect, adiabaticity for $H_{x}$ implies that

\begin{equation}
\int_{0}^{s} \overline{K}_{x} (\sigma) d\sigma \rightarrow O(1/\tau)\;,
\label{fin1}
\end{equation}

\noindent
which, in turn, implies that the kernel cannot contain terms that depend linearly on $\tau$. This means that matrix elements $\langle \widetilde{E}_{m}^{a}(s)|\widetilde{H}(s,\tau)|\widetilde{E}_{n}^{a}(s)\rangle\approx 0$. Thus, $\widetilde{H}$ must be close to $\widetilde{H}_{a}$ or it must be a constant with small time-dependent terms that vary slowly. The first case leads to a contradiction, since then the arguments of the exponentials get canceled and the system will not be adiabatic. Therefore, $\widetilde{H}$ must be given by a nearly constant term. Hence, Eq. (\ref{Eq1x}) gives

\begin{equation}
\left|\frac{\langle E_{m}^{x}(t)|\dot{E}_{n}^{x}(t)\rangle}{E_{n}^{x}(t)-E_{m}^{x}(t)}\right| \approx \left|\frac{\langle \widetilde{E}_{m}^{a}(t)|\dot{\widetilde{E}}_{n}^{a}(t)\rangle}{\widetilde{E}_{m}^{a}(t)-\widetilde{E}_{n}^{a}(t)}\right|<<1\;, \forall n,m \;;\forall t \;,
\label{Eq2x}
\end{equation} 

\noindent
and no inconsistency results.

\subsubsection{$H_{x}$ contains strong resonant oscillatory terms}\label{cincod2}

Previous studies \cite{deny0, explanations} have related the Marzlin-Sanders inconsistency to the existence of resonances but have not discriminated between weak and strong perturbations. These previous works did not consider either the $\tau \rightarrow \infty$ limit. Thus, the full extent of the inconsistency and its relation to the standard proof of the adiabatic theorem was not analyzed before. 

It is well known that the existence of exact resonances precludes adiabatic behavior \cite{comparat, deny0, explanations, schiff}.  Thus, we have

\begin{equation}
\int_{0}^{s} \overline{K}_{x} (\sigma) d\sigma \nrightarrow O(1/\tau)\;.
\label{fin2}
\end{equation}

On the other hand, $U_{x}$ contains strong oscillatory nondiagonal terms, since it relates an adiabatic Hamiltonian to a Hamiltonian with strong oscillatory terms. Thus, $\widetilde{H}=i\hbar \dot{U}_{x}U_{x}^{\dagger}$ will contain strong oscillatory terms too, and will be very different from $H_{a}$. In spite of the fact that the arguments of the exponentials in the kernel,  Eq. (\ref{big}),  are not canceled, the kernel integral does not go to zero because the matrix elements  $\langle \widetilde{E}_{m}^{a}(s)|\widetilde{H}(s, \tau)|\widetilde{E}_{n}^{a}(s)\rangle$ are nonmonotonic functions that depend on $\tau$.

The question at stake here is if Eq. (\ref{Eq1}) holds under these circumstances. If it does, there will be an inconsistency. Note the difference with Tong's dual system, for which $U_{x}=U_{a}$, $\widetilde{H}=H_{a}$, and the matrix elements  $\langle \widetilde{E}_{m}^{a}(t)|\widetilde{H}|\widetilde{E}_{n}^{a}(t)\rangle=0$. Instead, when the resonances are strong these matrix elements will not be small. Then the matrix elements $\langle E_{m}^{x}(t)|\dot{E}_{n}^{x}(t)\rangle$ will not be small either. Therefore, the condition, Eq. (\ref{Eq1}), will not be satisfied, and no inconsistency results.

\subsubsection{$H_{x}$ contains strong non-resonant oscillatory terms}\label{cincod3}

Nonresonant oscillatory terms can be safely averaged out, and therefore the exact time evolution is driven by the average Hamiltonian, which does not contain oscillating terms anymore.  Thus, in the $\tau \rightarrow \infty$ limit the average Hamiltonian will be adiabatic. Numerous cases exist in the bibliography showing the adequacy of this approach. A well-known example is the time evolution of the molecular alignment that takes place when a molecule interacts via its polarizability with a strong nonresonant laser pulse. This interaction depends on the square of the electric field. The time evolution of the rotational wave functions can be faithfully studied by taking into account only the laser envelope after averaging out the rapid oscillatory terms $\cos^{2}(\omega t)$, where $\omega$ is the nonresonant laser frequency \cite{ortigoso}. Thus, if the laser pulse is long enough the evolution is adiabatic, and Sec. \ref{cincod1} applies. 

Summarizing, the Marzlin-Sanders inconsistency takes place for a system $S_{x}$ {\it only} if the kernel of the Volterra evolution equation does not oscillate at frequencies that increase with $\tau$ and does not contain terms that vary nonmonotonically. The kernel, Eq. (\ref{big}), fulfills these two conditions if $\widetilde{H}\pm H_{a}=0$. The case $\widetilde{H}=H_{a}$ implies, from 
Eq. (\ref{generalU}), $U_{x}=U_{a}$, and the resulting $H_{x}$ is precisely the dual Hamiltonian used by Tong \cite{tong} (i.e., $H_{x}=H_{b}=-U_{a}^{\dagger}H_{a}U_{a}$). Contrarily, for the positive sign choice in the transformation that relates $H_{x}$ and $H_{a}$, the kernel satisfies the two conditions if $\widetilde{H}=-H_{a}$. 
In this case, $U_{x}$ is the exact evolution operator ($\approx U_{A}^{*}$) for the reversed Hamiltonian $-H_{a}$.  In rigour, the kernel fulfills the two cited conditions not only if $\widetilde{H}=\pm H_{a}$ but also if $\widetilde{H}$ admits a full asymptotic expansion, in powers of $\tau^{-1}$, in which the zero-order term is given by $\pm H_{a}$ [i.e., if $\widetilde{H}(s, \tau)=\pm H_{a}(s)+\sum_{j}(i\hbar/\tau)^{j} H^{j}(s)$], so the  properties derived in Sec. \ref{tres} are approximately valid. 

\section{Discussion}\label{seis}

The so-called Marzlin-Sanders inconsistency can be described as follows. If Eq. (\ref{Eq1}) is satisfied for a Hamiltonian $H_{a}$ for which the adiabatic approximation holds, it will be satisfied too for the Hamiltonian $H_{b}=-U_{a}^{\dagger}H_{a}U_{a}$, where $U_{a}$ is the exact evolution operator for $H_{a}$. However, in general, $H_{b}$ does not satisfy the adiabatic approximation \cite{tong}. This implies that Eq. (\ref{Eq1}) is not a sufficient condition for a system to hold the adiabatic approximation. 

The full extent of the inconsistency has been much debated and doubts about the consistency of the own adiabatic theorem have been raised. Here, we have proven that these doubts cannot be sustained.
Given a Hamiltonian $H_{a}(s=t/\tau)$ for which an adiabatic theorem holds in the $\tau \rightarrow \infty$ limit, the dual Hamiltonian $H_{b}(s, \tau)=-U_{a}^{\dagger}(s,\tau)H_{a}(s)U_{a}(s,\tau)$ does not satisfy the theorem because $H_{b}$ contains resonant terms that oscillate infinitely fast although their amplitude go asymptotically to zero. Due to these terms the dual Hamiltonian does not vary infinitely slowly when $\tau\rightarrow \infty$. Also, the derivatives of the eigenprojectors are not defined. Thus, no inconsistency affects the quantum adiabatic theorem because the dual Hamiltonian does not satisfy key premises required in the proofs of the theorem. 

In Sec. \ref{dosa}, we showed that premises (p1)-(p3) were imposed to guarantee that some oscillatory integrals that arise in the evolution equations, due to the existence of spectral gaps, go to zero in the adiabatic limit. Therefore, they only make strict sense in such a limit and cannot be used to determine the validity of the adiabatic approximation. However, the adiabatic approximation does not hold for the dual system because the same integrals that do not go to zero in the adiabatic limit are not small for finite $\tau$.
On the other hand, the conditions Eq. (\ref{Eq1}) hold for the dual Hamiltonian because the matrix elements $\langle E_{m}^{b}(s)|\dot{E}_{n}^{b}(s) \rangle$, although they oscillate fast, have a small magnitude.

Systems affected by the inconsistency can be easily identified because the instantaneous eigenprojectors of the dual Hamiltonian oscillate with negligible amplitude, with respect to a baseline, at frequencies resonant with the energy levels of the system. This baseline is defined by the eigenprojectors at the initial time. As a consequence, the instantaneous eigenprojectors change very little with time. However, the condition, Eq. (\ref{adiacon2}), is necessary but not sufficient for the inconsistency to occur, because it is equally satisfied by the Hamiltonian $-H_{b}$, which is adiabatic. 

On the other hand, the presence of fast oscillations of very small amplitude does not invalidate the adiabatic approximation unless they cancel the oscillatory factors arising from the gap condition (i.e., unless they are resonant). Thus, resonant terms, even if very small, make the gap condition irrelevant.

Also, we have proven, by using an approach different to that of Comparat \cite{comparat}, that the adiabatic approximation does not hold for Hamiltonians with strong oscillatory terms (resonant or nonresonant). In addition, our analysis indicates that, in this case, Eq. (\ref{Eq1}) is not satisfied. Hence, we conclude that Eq. (\ref{Eq1}) is a necessary and sufficient condition for a system to satisfy the adiabatic approximation unless the time-dependent part of the Hamiltonian contains resonant terms of very small magnitude. 

In cases for which an analytical expression is known for a particular Hamiltonian, it should be clear, due to the obvious presence of oscillatory terms, that the system does not satisfy the adiabatic approximation even if Eq.(\ref{Eq1}) holds. Contrarily, when no such analytical expression is known, the inconsistency may be more relevant. This case occurs when the Hamiltonian is known only through experimental information on its energy levels and eigenstates. 

In conclusion, the present work eliminates the mystery that has surrounded the adiabatic theorem since the publication of Ref. \cite{marzlin}.  Although it is expected that only a few systems will be affected by the inconsistency the condition Eqs. (\ref{condition1}), (\ref{adiacon2}) must supplement the quantitative conditions Eq. (\ref{Eq1}) whenever the validity of the adiabatic theorem and/or adiabatic approximation are investigated.

\section*{Acknowledgments}

I am grateful to Pablo Yubero for helpful discussions. Financial support from the Spanish Government through the MICINN (project FIS2010-18799) is acknowledged.

\end{document}